\documentclass[twocolumn,aps,floatfix]{revtex4}
\usepackage{amssymb}
\usepackage{graphicx}
\usepackage{amsmath}
\usepackage[T1]{fontenc}
\usepackage{pstricks}

\begin{document}
\title{Resonant dynamics of chromium condensates}

\author{Tomasz \'Swis{\l}ocki,$^1$ Jaros{\l}aw H. Bauer,$^2$ Mariusz Gajda,$^{1,3}$ and Miros{\l}aw Brewczyk$^{3,4}$}

\affiliation{
\mbox{$^1$Institute of Physics PAN, Al. Lotnik\'ow 32/46, 02-668 Warsaw, Poland}  \\
\mbox{$^2$Wydzia{\l} Fizyki i Informatyki Stosowanej, Uniwersytet {\L}\'odzki,
         ul. Pomorska 149/153, 90-236 {\L}\'od\'z, Poland }  \\
\mbox{$^3$Center for Theoretical Physics PAN, Al. Lotnik\'ow 32/46, 02-668 Warsaw, Poland} \\
\mbox{$^4$Wydzia{\l} Fizyki, Uniwersytet w Bia{\l}ymstoku, 
                             ul. Lipowa 41, 15-424 Bia{\l}ystok, Poland}    }

\date{\today}

\begin{abstract}

We numerically study the dynamics of a spinor chromium condensate in low magnetic fields. We show that the condensate evolution has a resonant character revealing rich structure of resonances similar to that already discussed in the case of alkali-atoms condensates. This indicates that dipolar resonances occur commonly in the systems of cold atoms. In fact, they have been already observed experimentally. We further simulate two recent experiments with chromium condensates, in which the threshold in spin relaxation  and the spontaneous demagnetization  phenomena were observed. We demonstrate that both these effects originate in resonant dynamics of chromium condensate.
\end{abstract}

\maketitle

\section{Introduction}
\label{introduction}

Experimental achievement of chromium condensates \cite{52Cr1,52Cr2} has started a huge interest in dipolar ultracold gases \cite{Lewenstein_review,Ueda_review}. First experiments with chromium condensates already showed spectacular features uncovered during expansion \cite{Pfau_expansion} and collapse \cite{Pfau_collapse} of atomic cloud and related to the anisotropy and long range character of dipole-dipole interactions. They were not present for alkali-atoms condensates. Another aspect of dipolar interactions in chromium is associated with its strength, which is two orders of magnitude larger than for alkalis. Frequently a technique based on Feshbach resonances is used to even enhance dipolar effects in a condensate \cite{Pfau_expansion}. In such a way the experimental study of the properties of almost purely dipolar systems becomes possible. Recently, condensates of atoms possessing even larger than chromium magnetic moments have been obtained experimentally. These are the condensates of dysprosium \cite{Dy} and erbium \cite{Er} atoms.

Magnetic dipolar interactions exhibit remarkable symmetries -- they couple atomic spin with orbital degrees of freedom. As a result coherent transfer of atoms between neighboring Zeeman states is allowed. In fact, only dipolar interactions can trigger the nontrivial spin dynamics when all atoms are initially put in a state with maximal spin projection. Transmitting spin into orbital motion is the essence of the Einstein-de Haas effect \cite{EdH} and was already numerically studied for both chromium \cite{Ueda_1,Santos} and rubidium \cite{KG_1,Swislocki_1,KG_2} condensates. In the latter case the dipolar resonances have been found out. Only in narrow intervals of values of external magnetic field a nonzero transfer of atoms between Zeeman components is possible. This happens because orbital motion of atoms is quantized and a particular amount of energy is necessary to promote atoms to higher energy orbital state. This required energy is just the Zeeman energy. In this paper we show that dipolar resonances are present also in chromium condensates. Indeed, they have been just observed experimentally \cite{Laburthe_3} but signatures of their existence can be already found in the results of earlier experiments. For example, as we show in this article, the threshold in spin relaxation observed in experiment of Ref. \cite{Laburthe_1} appears just as a front of a group of overlapped dipolar resonances. 

Contact interactions, on the other hand, preserve projection of total spin of colliding atoms which means that the magnetization of the system remains constant during the evolution. Contrary to the case of alkalis the chromium atoms have large spin-dependent contact interactions. In low enough magnetic fields they can overwhelm the Zeeman energy and many of new quantum phases with atoms distributed over a few components are possible \cite{Santos,Ho}. Dipolar interactions do not conserve the magnetization and hence they help to reach all these phases while initially chromium condensate is in the ferromagnetic state. Therefore, the dipolar interactions can lead to the demagnetization of the system. Indeed, an experiment with chromium condensate has been performed \cite{Laburthe_2} showing that in ultralow magnetic fields chromium gas looses its magnetization by spreading atoms initially populating a single state to all possible states.

The paper is organized as follows. In Sec. \ref{system} we describe the equations that govern the evolution of the chromium spinor condensate. Then in Sec. \ref{resonant} numerical results regarding the resonant dynamics of chromium condensate are presented. Secs. \ref{relaxation} and \ref{demagnetization} discuss experimental observations reported in Refs. \cite{Laburthe_1} (presence of a threshold in spin relaxation) and \cite{Laburthe_2} (spontaneous demagnetization at extremely low magnetic fields), respectively and show that both of them originate from the resonant character of dipolar interactions. We end with conclusions in Sec. \ref{summary}.

\section{Description of the system}
\label{system}

We study a chromium spinor condensate within the mean-field approximation. The wave function $\psi ({\bf r})=(\psi_3({\bf r}), \psi_2({\bf r}), \psi_1({\bf r}), \psi_0({\bf r}), \psi_{-1}({\bf r}), \psi_{-2}({\bf r}), \psi_{-3}({\bf r}))^T$ of the system fulfills the following equation
\begin{eqnarray}
i\hbar \frac{\partial}{\partial t}\,  \psi ({\bf r}) = ({\cal{H}}_{sp} + {\cal{H}}_c + {\cal{H}}_d )\, \psi ({\bf r})  \,,
\label{eqmot}
\end{eqnarray}
where the effective Hamiltonian is written as a sum of three terms. The first part, ${\cal{H}}_{sp}$, is diagonal in Zeeman components and represents the single-particle contribution ${\cal{H}}_{sp}=-\frac{\hbar^2}{2 m} \nabla^2 + V_{ext}({\bf r}) -\boldsymbol{\mu} {\bf B}$. It includes the kinetic, potential, and Zeeman energies, respectively. The external magnetic field ${\bf B}$ is aligned along the $z$ axis and the magnetic moment operator of chromium atom is $\boldsymbol{\mu}=g_L \mu_B {\bf s}$, where $\mu_B$ is the Bohr magneton, $g_L=-2$ is the Lande factor, and ${\bf s}$ are standard $7\times7$ spin matrices (the spin quantum number $s=3$ for the chromium atom in the ground state). 

Two other terms in (\ref{eqmot}) originate from the two-particle interactions. The second one, ${\cal{H}}_c$, corresponds to the contact interactions and can be written in the form
\begin{equation}
{\cal{H}}_{c} ({\bf r}) = \int d^{\,3}r'\, \psi^{\dagger}({\bf r'}) \, V_c ({\bf r} - {\bf r'})\, \psi({\bf r'}) \,.
\label{contact}
\end{equation}
Since the contact interactions operator $V_c ({\bf r}, {\bf r'})$ commutes with the total spin of colliding atoms as well as with its projection on the $z$ axis, it can be spectrally decomposed as
\begin{equation}
V_c ({\bf r}, {\bf r'}) = \delta ({\bf r} - {\bf r'}) \sum_{S=0}^{2s} g_S \sum_{M=-S}^S |S M\rangle \langle S M|.
\label{Vc}
\end{equation}
Here, $|S M\rangle$ is a state of a pair of atoms with the total spin $S$ and projection $M$, which in turn is expanded in a product two-particle basis
\begin{equation}
|S M\rangle = \sum_{m_1,m_2} (s,m_1;s,m_2|S M) \; |s,m_1\rangle |s,m_2\rangle  \,,
\label{SM}
\end{equation}
where the symbol $(s,m_1;s,m_2|S M)$ is the Clebsch-Gordan coefficient and $|s,m\rangle$ is the state of a single atom with the spin $s$ and projection $m$. The strength of the contact interactions is characterized by the $s$-wave scattering length $a_S$ for the total spin $S$ of colliding atoms as $g_S=4\pi\hbar^2a_S/m$. Due to the symmetry of the wave function of a pair of bosonic atoms only $S=0,2,4,6$ are allowed. Using (\ref{contact}), (\ref{Vc}), and (\ref{SM}) all together one gets
\begin{eqnarray}
({\cal{H}}_{c})_{m_1 m_1'} &=& \sum_{S=0}^{2s} g_S \sum_{M=-S}^S  \sum_{m_2,m_2'}  (s,m_1;s,m_2|S M)
\nonumber  \\
&\times& (SM|s,m_1';s,m_2') \, \psi^{*}_{m_2}({\bf r})  \, \psi_{m_2'}({\bf r})  \,,
\label{Hcab}
\end{eqnarray}
where $m_1,m_1'=3,2,...,-3$.

Finally, the third term in (\ref{eqmot}) is related to the dipolar interactions and is written as
\begin{equation}
{\cal{H}}_{d} ({\bf r}) = \int d^{\,3}r'\, \psi^{\dagger}({\bf r'}) \, V_d ({\bf r} - {\bf r'})\, \psi({\bf r'}) \,,
\label{dipolar}
\end{equation}
where the interaction energy of two atoms with magnetic moments $\boldsymbol{\mu}_1$ and $\boldsymbol{\mu}_2$, positioned at ${\bf r}$ and ${\bf r'}$ is
\begin{eqnarray}
V_d ({\bf r}, {\bf r'}) = \frac{\boldsymbol{\mu}_1 \, \boldsymbol{\mu}_2}{|\bf r-\bf r'|^3}-3
\frac{
[\boldsymbol{\mu}_1 \, (\bf r-\bf r')] \,
[\boldsymbol{\mu}_2 \, (\bf r-\bf r')]}
{|\bf r-\bf r'|^5}   \,.
\label{ddi}
\end{eqnarray}
The dipolar energy can be expanded in a basis of spherical harmonics \cite{PethickSmith}
\begin{equation}
V_d \propto 
\sum_{\lambda=-2}^{2} Y_{2\lambda}^{*}({\hat{\bf r}})\Sigma_{2\lambda}   \;,
\label{expansion}
\end{equation}
where only rank-2 spherical harmonics contribute. $\hat{\bf r}$ denotes a unit vector in the direction of relative position of two atoms. A rank-2 spherical tensor $\Sigma_{2\mu}$ is defined as
\begin{eqnarray}
\Sigma_{2,0} =&& -\sqrt{\frac{3}{2}}(s_{1z} s_{2z}-{\bf s}_1 \cdot {\bf s}_2 /3)    
\nonumber \\
\Sigma_{2,\pm1} =&& \pm\frac{1}{2} (s_{1z} s_{2\pm}+s_{1\pm}s_{2z})  
\nonumber \\
\Sigma_{2,\pm2} =&& -\frac{1}{2} s_{1\pm} s_{2\pm}    \,,
\label{tensor}
\end{eqnarray}
where $s_{1\pm}$ and $s_{2\pm}$ are raising and lowering atomic spin projection operators. It is clear from (\ref{tensor}) that the spin projection of a pair of colliding atoms can change at most by $2$, however, the spin projection of a single atom changes maximally by $1$. Therefore the matrix ${\cal{H}}_{d}$ gets tridiagonal. It is easy to check that diagonal elements are $({\cal{H}}_{d})_{\alpha\alpha}=\alpha ({\cal{H}}_{d})_{11}$ ($\alpha=3,2,...,-3$) while detailed form of $({\cal{H}}_{d})_{11}$ is given in Appendix \ref{first}. The off-diagonal elements are $({\cal{H}}_{d})_{\alpha,\alpha-1}=\sqrt{(4-\alpha) (3+\alpha)/12}\,\, ({\cal{H}}_{d})_{10}$ ($\alpha=3,2,...,-2$), where again $({\cal{H}}_{d})_{10}$ is written in Appendix \ref{first}. Other elements of ${\cal{H}}_{d}$ matrix are found by using its hermiticity property.

\section{Resonant dynamics of chromium condensates}
\label{resonant}

We solve the set of Eqs. (\ref{eqmot}) assuming that initially all chromium atoms are in $m_s=-3$ Zeeman component. It is the ground state of a chromium condensate when external magnetic field is large enough \cite{Santos,Ho}. We suddenly reverse the direction of magnetic field and change its value. Experimentally, usually instead of changing the direction of magnetic field the atoms are transferred (for instance, by radio-frequency sweep \cite{Laburthe_1}) to the $m_s=+3$ Zeeman state and then the field is decreased. In such a way the dipolar relaxation to other Zeeman components becomes energetically allowed and their populations as a function of magnetic field can be investigated.

\begin{figure}[thb]
\resizebox{3.3in}{2.0in}{\includegraphics{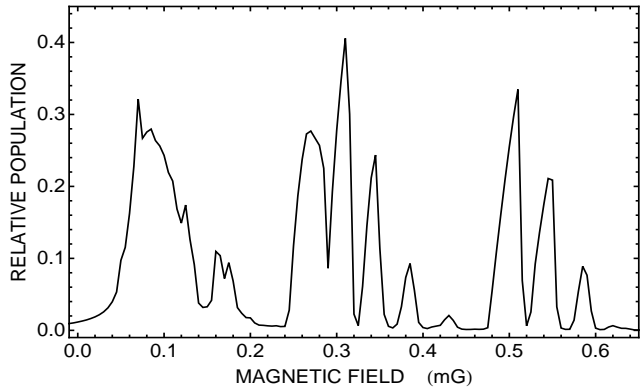}}
\caption{Relative population of $m_s=-2$ component (calculated by using the maximal transfer within $20$ms) of a spinor chromium condensate as a function of magnetic field. The frequencies of an axially symmetric cigar-shaped trap are $\omega_{x,y}=2\pi \times 400$Hz and $\omega_z=2\pi \times 100$Hz. The initial number of atoms in $m_s=-3$ state is $N_{-3}=10^4$ which corresponds to the density at the trap center equal to $1.5\times 10^{14}$cm$^{-3}$. First three groups of resonances are clearly visible. }
\label{res}
\end{figure}

\begin{figure}[thb] 
{\includegraphics[width=8cm]{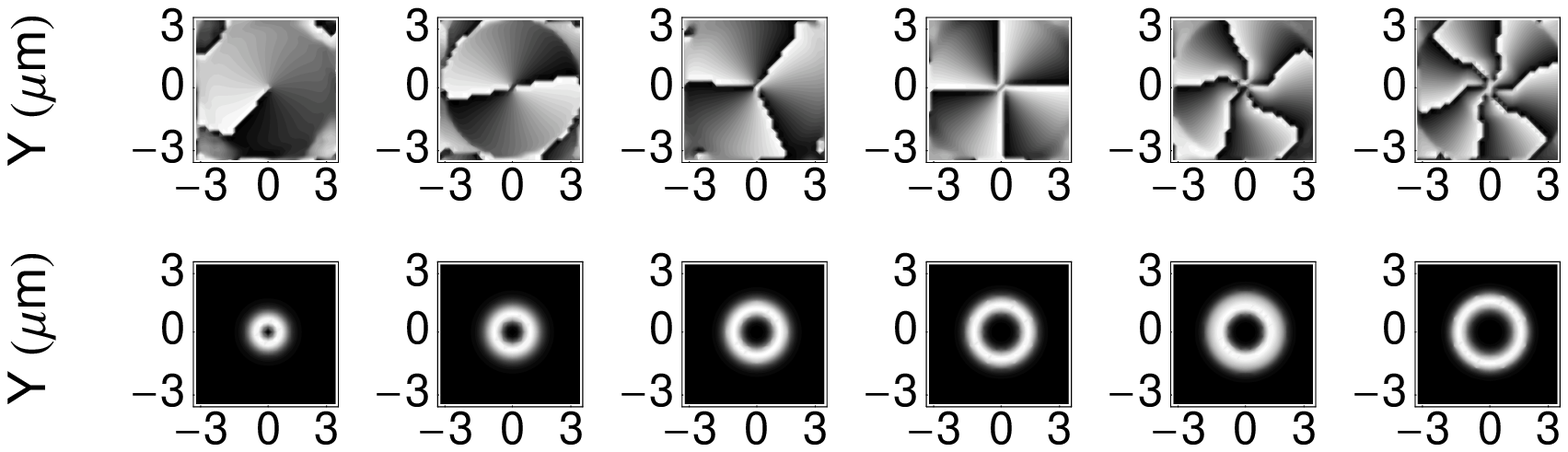} \\
\includegraphics[width=8.7cm]{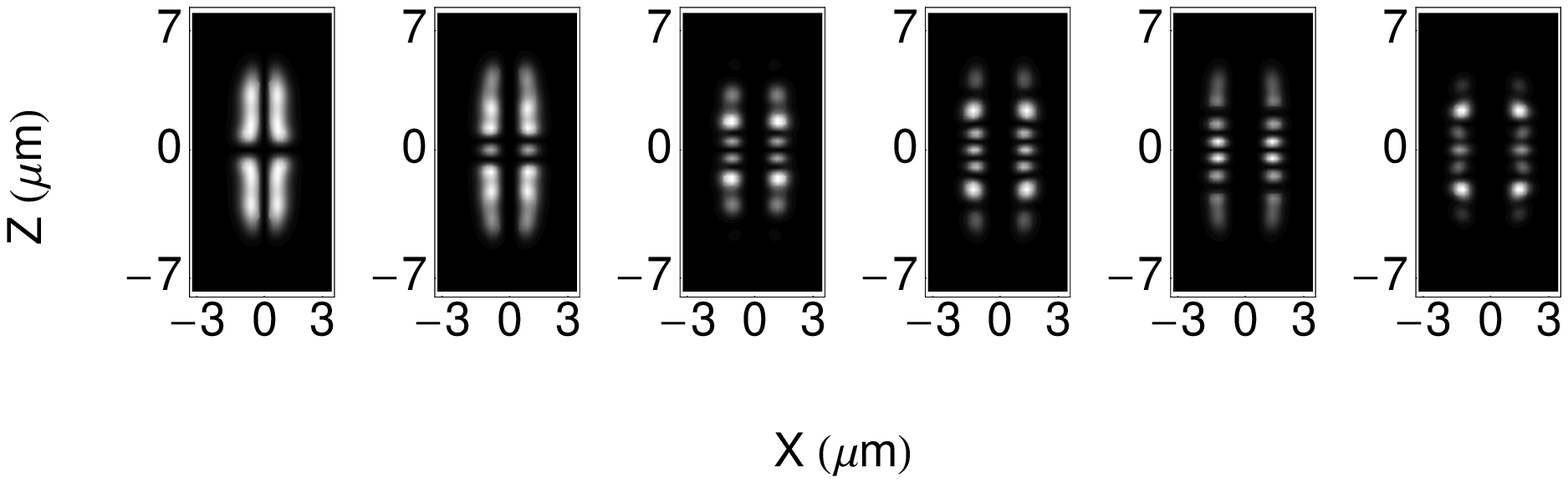}}
\caption{Phase (upper panel) and density (middle and lower panels) of spinor components for the magnetic field $B_{res}=0.07$mG (corresponding to the first peak resonance in Fig. \ref{res}) after $10$ms of evolution. The upper and middle panels show the phases and the densities in the $xy$ plane for a particular value of position in $z$ direction whereas the lower one exhibits the densities in the $xz$ plane. Spinor components are ordered in columns (from $m_s=-2$ to $m_s=+3$). }
\label{dens}
\end{figure}

\begin{figure}[thb] \resizebox{3.6in}{1.5in}
{\includegraphics{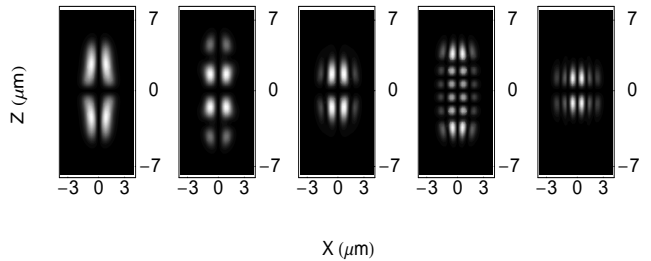}}
\caption{Densities (in the $xz$ plane) of $m_s=-2$ spinor component characteristic for resonances visible in Fig. \ref{res}. The first, third, and fifth frames show density patterns for main resonances whereas the second and the fourth describe side peaks (the first two). The main resonances are related to the successive radial excitation of the atomic cloud. On the other hand, the side resonances correspond to axial excitations.  }
\label{densres}
\end{figure}

First, we consider a chromium condensate at low densities. The initial number of chromium atoms is $N_{-3}=10^4$. The atoms are confined in an axially symmetric cigar-shaped trap with frequencies $\omega_{x,y}=2\pi \times 400$Hz and $\omega_z=2\pi \times 100$Hz which results in a central atomic density of about $10^{14}$cm$^{-3}$. In Fig. \ref{res} we plot relative population of $m_s=-2$ state for magnetic fields below $1$mG at the moment of maximal transfer of atoms. Clearly, three groups of resonances are visible. All resonant peaks can be identified by looking at the spatial dependence of $\psi_{-2}({\bf r})$ component as demonstrated in Figs. \ref{dens} and \ref{densres}. The first column in Fig. \ref{dens} proves that for magnetic field $B_{res}=0.07$mG atoms going to $m_s=-2$ Zeeman state acquire one quanta of rotational motion (in the $xy$ plane) and one quanta of excitation in $z$ direction (since two rings in $z$ direction are visible). The excitation energy of such a single-particle state equals $\hbar \omega_{\perp} + \hbar \omega_z$ which, in units of milligauss, is $0.178$mG. This value is comparable to the value of $0.07$mG found numerically (see Fig. \ref{res}, the first peak in the first group of resonances) and the difference is due to the presence of contact interactions. Contact interactions increase the mean field energy of the initial state, therefore less Zeeman energy is needed to satisfy the energy conservation responsible for appearance of the resonance.

Due to the energy conservation and weakness of the dipole-dipole interactions, the dynamics of the atomic cloud  with a free magnetization allows, to some extend, for mapping of the energy levels of the system onto efficiency of a transfer of atoms to the $m_s=-2$ component. At higher magnetic fields the transitions to other, more excited states become energetically allowed. Due to the cigar-shaped geometry assumed here, the resonances observed in Fig. \ref{res} are grouped according to the number of excitation quanta in the radial direction, $(2 n_{\perp}+1) \hbar \omega_{\perp}$, where $n_{\perp}$ is an integer.  One can see three such groups in Fig. \ref{res} corresponding to $n_{\perp}=0,1,2$. On top of every such an excitation there are softer axial excitations $ n_{z} \hbar \omega_z$, with $n_{z}$ being odd. This results in the fine structure within each group.
For example, at magnetic fields around $0.3$mG the second group of resonances is visible (Fig. \ref{res}) with the excitation energy equal to $3 \hbar \omega_{\perp} + \hbar \omega_z$ assuming no contact interactions are present (this energy is, in fact, related to the most left peak of the group). The third frame in Fig. \ref{densres} proves that indeed the orbital state becomes radially excited. Similarly, the last group of peaks in Fig. \ref{res} corresponds to the spatial states which acquire two quanta of radial excitation (as confirmed by the last frame in Fig. \ref{densres}). As already mentioned, within each group of resonances there are peaks related to axial excitations. Such excitations are energetically less demanding since the axial trap frequency is much less than the radial one. For example, the second peak in the first group and the third one in the second group in Fig. \ref{res} originate (as confirmed by the second and fourth frames of Fig. \ref{densres}) from the transfer of atoms to axially excited states with three and five quanta of axial excitation, respectively.

\begin{figure}[thb]
\resizebox{3.3in}{2.0in}{\includegraphics{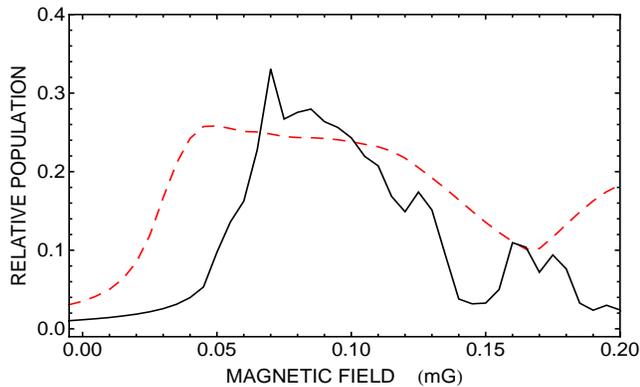}}
\caption{(color online). Comparison of resonance structure for condensates with initial number of atoms $N_{-3}=10^4$ as in Fig. \ref{res} (solid line) and for $N_{-3}=5\times10^4$ (dashed line). In the latter case the peak density is twice larger ($2.8\times 10^{14}$cm$^{-3}$). The shift towards lower magnetic fields, stronger overlap,  and larger extend over the region corresponding to opposite direction of magnetic field are clearly visible. }
\label{shift}
\end{figure}

Finally, we compare the resonance structure for chromium condensates with different number of atoms. The main observation is that by increasing the atomic density (i.e. by increasing the number of atoms in a condensate) resonances become stronger overlapped merging almost first two groups of resonances visible in Fig. \ref{res} (see Fig. \ref{shift}). There appears also a shift towards smaller magnetic fields. Evidently, the shift of resonances is related to contact interactions which get more pronounced for higher atomic density. Its appearance is visible already on a level of first order perturbation calculus since the scattering length $a_6$ is positive. For large enough atomic densities the main resonance extends even up to the region where magnetic field is of the same direction as the field used to prepare the initial sample of atomic cloud. This feature is responsible for the occurrence of demagnetization of initially polarized chromium condensate as discussed in Sec. \ref{demagnetization}.

\section{Dipolar relaxation in optical lattice}
\label{relaxation}

In experiment of Ref. \cite{Laburthe_1} the chromium condensate prepared in its ground state $m_s=-3$ is adiabatically loaded in the lowest energy band of two-dimensional optical lattice. Then, with the help of radio-frequency sweep the chromium atoms are transferred to the most upper state $m_s=+3$. After that the dipolar relaxation of atoms to other Zeeman components becomes energetically allowed. What is observed in experiment is a threshold in dipolar relaxation as a function of the magnetic field. Below the threshold no transfer of atoms to other components is present. Above the threshold (i.e. for high enough magnetic fields) the relaxation to $m_s=2$ Zeeman state is observed. For the lattice with maximal depth in each direction of about $25 E_R$, where $E_R$ is the recoil energy, the threshold magnetic field is approximately at $42$mG. Above threshold, as it is said in Ref. \cite{Laburthe_1}, first excited band in the lattice becomes populated by atoms in $m_s=2$ Zeeman state. No population in the second band is reported. Also no production of rotating states in each lattice site is observed as could be according to the Einstein-de Haas effect. This result is explained in terms of decaying process of a quantized vortex due to the tunneling.

\begin{figure}[thb] 
\resizebox{3.3in}{2.0in}{\includegraphics{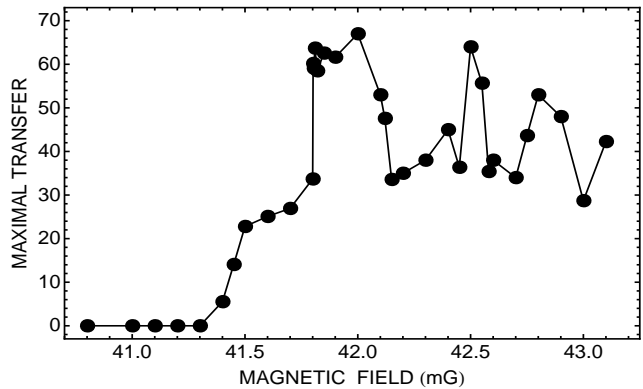}}
\caption{Maximal transfer to $m_s=-2$ component within first $30$ms of evolution for various external magnetic fields. $N=160$ atoms being initially in the state  $m_s=-3$ are trapped in a square plaquette of four optical sites. The plaquette is formed by two counter-propagating pairs of laser beams (with the wavelengths equal to $532$nm) positioned in $xy$ plane and the harmonic trapping potential corresponding to $2\pi \times 400$Hz frequency in $z$ direction. Clearly, the threshold around $42$mG and wide for $0.6$mG  is visible. }
\label{threshold}
\end{figure}

To verify the existence of the threshold in dipolar relaxation with respect to the magnetic field we performed numerical simulations on a square plaquette of four optical sites. The number of chromium atoms per site is $40$ and the peak density is of the order of $10^{15}$cm$^{-3}$. The plaquette is created by two counter-propagating pairs of laser beams with the wavelengths equal to $532$nm positioned in the $xy$ plane and a harmonic trapping potential of $2\pi \times 400$Hz frequency in $z$ direction. The depth of the plaquette in each direction is $25 E_R$. Each site potential can be approximated by a harmonic trap with the frequency equal to $\omega_\perp=2\pi \times 136$kHz and a simple estimation neglecting contact interactions gives the position of the first dipolar resonance at $\hbar \omega_\perp /2 \mu_B=48$mG. Contact interactions move the position of this resonance which is then at the value of about $42$mG (approximately $120$kHz in units of frequency), see Fig. \ref{threshold}. The value of this threshold remains in agreement with experimental results of Ref. \cite{Laburthe_1} (see Fig. $2$, where the population of the first excited band after $25$ms of dipolar relaxation is plotted).

The resonance described above is associated with the absorption of one quanta of rotational energy in the $xy$ plane. However, as it was discussed in Ref. \cite{Swislocki_1} for rubidium condensate, for large occupation of initial state the process during which one atom flips its spin dominates upon the dipolar collision. This is because the dipolar coupling term (see Eq.(\ref{dipolar})) responsible for this process is proportional to the density of the highly occupied $m_s=-3$ component, $|\psi_{-3}({\bf r})|^2$, i.e. the initial atomic density. On the contrary, the dipolar term which is responsible for simultaneous flip of spin of two interacting atoms is govern by a term proportional to $\psi^*_{-2}({\bf r})\psi_{-3}({\bf r})$. In this process the second excited band becomes populated and doubly charged vortices should appear together with even number of excitation quanta in axial direction. Note that initially there are no atoms in the $m_s=-2$ component, i.e. $\psi^*_{-2}({\bf r})=0$. Therefore, in the mean field limit this process can start only if some initial population is   present in $m_s=-2$ Zeeman state. In order to account for this process we introduced some small initial seed in this state. Nevertheless in our simulations we did not observe any significant transfer of atoms to the second excited band.

\begin{figure}[thb] 
{\includegraphics[width=8.5cm]{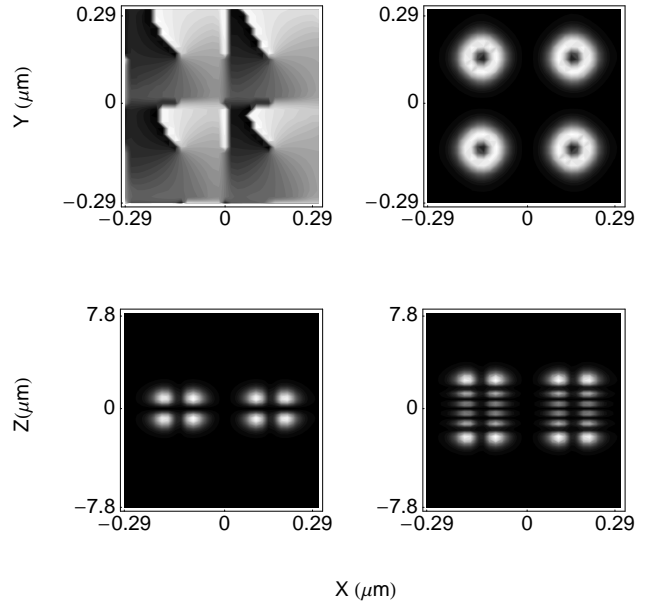}}
\caption{Upper panel: Typical phase (left frame) and density (right frame) of $m_s=-2$ atoms in the $xy$ plane for magnetic fields above the threshold. Lower panel: Densities of $m_s=-2$ component in the $xz$ plane for magnetic fields $41.85$mG (left frame) and $42.3$mG (right frame), respectively (lower panel). }
\label{dens2x2}
\end{figure}

Symmetries of dipolar interactions  have important consequences on the final states of $m_s=-2$ atoms. Namely if only one of the two interacting atoms flips its spin these final states must correspond to odd number of quanta of axial excitations. Consequently the spatial density must exhibit even number of rings in $z$ direction. Indeed, it is the case as demonstrated in Fig. \ref{dens2x2}. The upper panel shows typical phase and density of spatial states of $m_s=-2$ atoms during the evolution. Only singly quantized vortices are generated in plaquette sites (as opposed to what is suggested in Ref. \cite{Laburthe_1}). The lower panel, on the other hand, depicts that spatial structure of $m_s=-2$ atoms state gets more complicated while increasing the magnetic field. The number of rings in $z$ direction in each site increases. This behavior is like the one appearing in the case of a condensate confined in a single harmonic trap discussed in the previous section although no separate resonant peaks are now visible (see Fig. \ref{threshold}). This is because the resonances are broadened due to the presence of other plaquette sites. Simple estimation gives $4 \hbar \omega_z /2 \mu_B=0.57$mG as the value the magnetic field need to be increased to change the spatial structure of the state from the one with two rings (left frame) to that with six rings (right frame). This estimation remains in agreement with numerical calculations.

Although numerically found position of the threshold compares with experimental value reported in Ref. \cite{Laburthe_1} the width of  numerical threshold remains much narrower. This width should be related to the width of the excited lattice band to which atoms are promoted via dipolar interactions. Since in our case the spatial state of $m_s=-2$ atoms exhibits always a singly quantized vortex (see Fig. \ref{dens2x2}) only a first excited band (or rather quasiband since we work with $2\times2$ plaquette) is populated. The tunneling energy for the first excited band for the lattice with the lattice period equal to ($532/2$)nm and with the maximal depth of $25E_R$ is $J_1=2.98\times 10^{-2}E_R$. Therefore the width of the first excited band equals $4J_1=1.6$kHz which, in units of magnetic field, is $0.58$mG. Numerically obtained width of the threshold is just about this value (see Fig. \ref{threshold}).

Hence, our explanation of the appearance of the threshold differs from that presented in Ref. \cite{Laburthe_1}, where it is argued that mainly the second excited band is populated via dipolar interactions. Subsequently, atoms in the second excited band rapidly loose their orbital angular momentum as a result of tunneling and then via collisional deexcitation with atoms in the ground band they start to populate the first excited band. Therefore population accumulates in the first excited band as observed in experiment. In our calculations, on the other hand, we see direct population of the first excited band. This is because the spatial state of $m_s=-2$ atoms exhibits only one excitation in the $xy$ plane (since it is a singly quantized vortex). In our explanation the width of the threshold is therefore related to the width of the first excited band. We observe in our calculations, however, that vortices disappear on a time scale of the order of milliseconds. This time scale agrees with the tunneling time in the first excited band. It is an indication that vortices decease because of tunneling.

\section{Demagnetization in a harmonic trap}
\label{demagnetization}

Finally, we consider the spontaneous demagnetization phenomenon observed experimentally in a chromium condensate in extremely low magnetic fields. In this experiment (see Ref. \cite{Laburthe_2}) the chromium condensate is produced in an optical dipole trap in $m_s=-3$ state. The large enough magnetic field (about $20$mG in this case) was turned on during evaporation. For such a value of magnetic field the system is in the ferromagnetic phase \cite{Santos,Ho}. Next, the magnetic field was reduced to the values below $1$mG at which, according to the zero temperature phase diagram for chromium \cite{Santos,Ho}, the system prefers staying in other than ferromagnetic phases. These are the polar or cycling phases or the phases in which all Zeeman components are occupied. Indeed, population of all states was experimentally observed both in the case of chromium condensate confined in a harmonic trap as well as in an optical lattice in which case the peak density was almost an order of magnitude larger than for harmonic trap. In the latter case the demagnetization occurs for larger magnetic fields.

\begin{figure}[thb]
\resizebox{3.3in}{2.0in}{\includegraphics{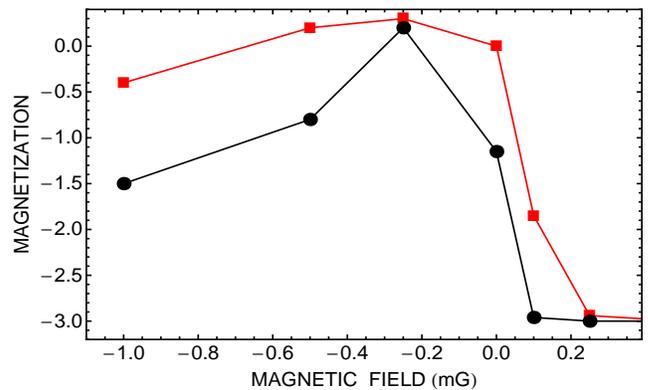}}
\caption{(color online). Magnetization of chromium condensate as a function of magnetic field for a condensate with $N_{-3}=2\times10^4$ atoms (peak density of $3.5\times 10^{14}$cm$^{-3}$, black bullets) and with $N_{-3}=10^5$ atoms (twice larger peak density, red squares). }
\label{magn}
\end{figure}

To study spontaneous demagnetization we did numerical calculations for a chromium condensate confined in a harmonic trap. Trap frequencies equal $(320,400,550)$Hz as in experiment \cite{Laburthe_2}, the number of atoms is $2\times10^4$. It gives the peak density as high as $3.5\times 10^{14}$cm$^{-3}$. The scattering lengths are: $a_6=102$, $a_4=63$, $a_2=-7$, and $a_0=91$ in units of the Bohr radius \cite{Laburthe_4,Ueda_1}. Initially, since external magnetic field is high enough, all atoms occupy $m_s=-3$ Zeeman state, i.e. the system is in ferromagnetic phase. We then suddenly decrease the magnetic field to the value below $1$mG and study succeeding evolution. It comes off the phase diagram at zero temperature that for magnetic fields below the threshold $B_{th}$ \cite{Ho}
\begin{equation}
2 \mu_B B_{th} = 0.68 \frac{2\pi \hbar^2 (a_6-a_4)}{m} n    \,,
\label{Bth}
\end{equation}
where $n$ is the atomic density, the system is no longer in ferromagnetic phase. Below $B_{th}$ the ground state of the system changes its character and some nontrivial dynamics is expected when the magnetic field is decreased from high to low values. In our case $B_{th}=0.21$mG which is of the same order of magnitude as we found numerically (equal approximately to $0.1$mG for the peak density of $3.5\times 10^{14}$cm$^{-3}$, see Fig. \ref{magn} ).

In Fig. \ref{magn} we plot the magnetization (defined as a projection of total spin, $\sum_{s=-3}^3 m_s N_s$, where $N_s$ is the number of atoms in $m_s$ component) of a chromium condensate as a function of magnetic field. We consider condensates with different number of atoms. The demagnetization begins for magnetic fields well below $1$mG in agreement with the formula (\ref{Bth}). According to this formula the onset of the demagnetization process occurs for higher magnetic fields when the condensate is denser. Fig. \ref{magn} shows that this feature is reproduced by numerics. It is also observed in experiment when the condensate is loaded into the optical lattice instead of the harmonic trap which results in the peak density almost an order of magnitude higher (see Fig. 2 in Ref. \cite{Laburthe_2}).

Reasoning based on the phase diagram alone is not able, however, to explain what is observed in numerics. It turns out that what happens strongly depends on the sign of the magnetic field. For negative magnetic fields (i.e. pointed towards the negative $z$ axis) we find even stronger depolarization, the magnetization of the system gets even smaller than for positive magnetic fields. The explanation is as follows. First, one has to remember that only dipolar interactions can trigger the process of populating the $m_s=-2$ state when all particles are initially polarized in $m_s=-3$ state. But as we already know from the previous sections, the transfer of atoms related to dipolar interactions is effective only on resonance. The first resonance occurs for low magnetic fields, mainly negative. When the density of a condensate increases the position of the first resonance is moved towards positive fields because of the contact interactions which shift the initial state energy towards the energy of $m_s=-2$ component. Therefore for dense enough condensates a significant transfer of atoms to $m_s=-2$ state becomes possible also for positive magnetic fields on the expense of the contact rather than  Zeeman energy. This is the origin of demagnetization mechanism as observed in experiment \cite{Laburthe_2}. However, we would like to emphasize that demagnetization process is actually related to the coupling mechanism which can trigger the transfer process. In the studied case the coupling is due to the dipole-dipole interactions which are effective only at the resonance. Depending on the density the first dipolar resonance can occur for positive or negative (low enough in both cases) magnetic fields. The  Eq. (\ref{Bth}) is therefore only a necessary condition for the demagnetization (energetic instability), but the dynamical transfer is related to the coupling mechanism which has a resonant character.

\begin{figure}[thb] 
\resizebox{3.4in}{2.1in}{\includegraphics{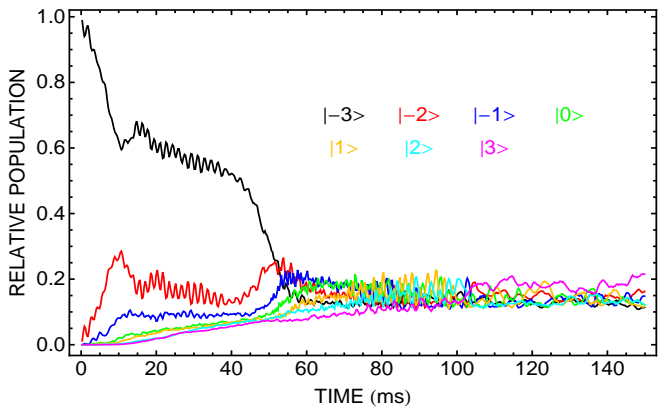}}
\caption{(color online). Relative populations of all Zeeman components as a function of time. Initial number of atoms in $m_s=-3$ state is $N_{-3}=2\times10^4$ and the value of the magnetic field equals $-0.25$mG (which after $150$ms of evolution results in the magnetization equal almost to zero, see Fig. \ref{magn}). After $60$ms of evolution atoms are already spread over all possible components.  }
\label{rel}
\end{figure}

\begin{figure}[thb]
\resizebox{3.3in}{2.1in}{\includegraphics{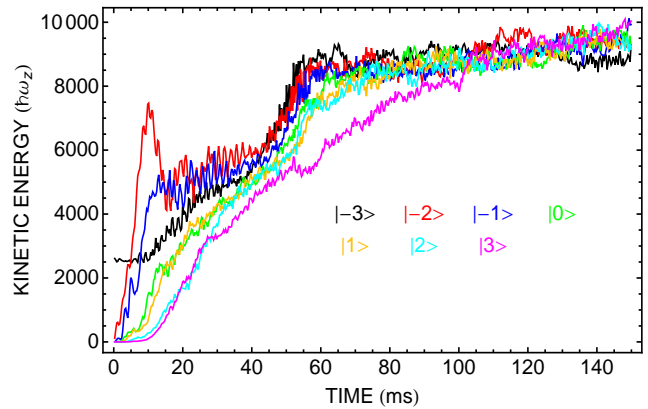}}
\caption{(color online). Kinetic energy of each component as a function of time for initially polarized condensate with $2\times10^4$ atoms in $m_s=-3$ Zeeman state placed in external magnetic field $B_{res}=-0.25$mG. Note that after $100$ms of evolution the kinetic energy in each component equalizes which strongly suggests that equilibration occurred in the system (see Ref. \cite{KG_3}).  }
\label{kinene}
\end{figure}

For magnetic field $B=-0.25$mG, for example, atoms get finally almost equally distributed over all Zeeman components, see Fig. \ref{rel}. In this case the magnetization is about zero. There are three distinct regimes during the demagnetization process. In the first one (of duration of a few milliseconds) the dynamics is mainly governed by the dipolar interactions. During the evolution the spatial states of $m_s=-2,-1,0,...$ components exhibit quantized vortices with charges equal to $-1,-2,-3,...$, respectively. After this short period, however, the contact interactions start to play dominant role and the thermalization process occurs - the circulation disappears. A signature of thermalization (the second regime) is already visible while looking at the kinetic energies of spinor components as a function of time (see Ref. \cite{KG_3}). Fig. \ref{kinene} shows that after $100\,$ms of evolution the kinetic energies of all components equilibrate. It can be checked within the classical field approximation \cite{CFA1,CFA2} that after $100\,$ms the fraction of total thermal atoms in each component becomes $1/7$. Hence, the kinetic energy per thermal atom gets the same in each Zeeman state which exactly means that the system has reached the third regime, the thermal equilibrium.

We have also checked whether the onset of demagnetization process is indeed related to the existence of the lowest energy dipolar resonance. One might assume that another scenario happens and the spontaneous demagnetization is triggered by the processes in which atoms are promoted to $m_s=-2$ state via dipolar interactions while their kinetic energy is changed into the Zeeman energy. No resonance condition is necessary for such a process. Therefore we added an extra noise to the initial state of atoms in $m_s=-3$ component which is equivalent to putting thermal atoms into the system \cite{CFA2} and repeated our calculations for different values of magnetic field. Although the initial energy of $m_s=-3$ component was increased by more than twice (resulting in its $30\%$ depletion) no significant change in populations and other properties was found. In particular, the magnetization curves (Fig. \ref{magn}) look the same. This result acts in favour of dipolar resonances based scenario.

\section{Summary}
\label{summary}
In conclusion, we have studied the dynamics of a spinor chromium condensate in low magnetic fields. We found out that the condensate dynamics has a resonant character. The appearance of dipolar resonances seems to be a common phenomenon in ultracold atomic gases in low magnetic fields. We show that already observed experimentally phenomena of appearance of a threshold in spin relaxation \cite{Laburthe_1} and of spontaneous demagnetization \cite{Laburthe_2}, in fact, both occur because of existence of dipolar resonances. In both cases the initial stage of the evolution is resonant which is so because of the presence of quantized vortices in populated Zeeman components. In the case of $2\times2$ plaquette we still observe vortices in $m_s=-2$ component in each plaquette site even though these sites are not axially symmetric. This is related to the fact that we have large number of atoms in each lattice site (see Ref. \cite{Pietraszewicz} for the discussion of the role of the trap anisotropy in the spin relaxation processes). The resonant evolution of a condensate is, however, terminated after a few milliseconds. On this time scale mechanisms like tunneling and thermalization, which destroy vortices, come into play. After long enough time (about $100\,$ms in the case of demagnetization process) spinor condensate reaches a thermal equilibrium.

\acknowledgments 

We are grateful to B. Laburthe-Tolra, P. Pedri, and T. Sowi\'nski for helpful discussions. The work was supported by the National Science Center grants No. DEC-2011/01/B/ST2/05125 (T.\'S., M.G.) and DEC-2012/04/A/ST2/00090 (M.B.).

\appendix
\section{Dipolar matrix elements}
\label{first}

The dipolar matrix element $({\cal{H}}_{d})_{11}$ equals
\begin{eqnarray}
&&({\cal{H}}_d (\mathbf{r}))_{11} = \gamma^2 \int d^3r' \left[
\frac{1}{|\mathbf{r}-\mathbf{r}'|^3}-3\frac{(z-z')^2}{|\mathbf{r}-\mathbf{r}'|^5}
\right] \nonumber \\
&&\times \sum_{m=-s}^s  m |\psi_m ({\bf r})|^2   \nonumber \\
&& -3 \gamma^2
\int d^3r' \frac{z-z'}{|\mathbf{r}-\mathbf{r}'|^5}[(x-x') - i (y-y')]  \nonumber \\
&&\times \sum_{m=-s+1}^s \sqrt{(4-m)(3+m)/4}\,\, \psi_m^{*} ({\bf r})\, \psi_{m-1} ({\bf r}) \nonumber \\
&& -3 \gamma^2
\int d^3r' \frac{z-z'}{|\mathbf{r}-\mathbf{r}'|^5}[(x-x') + i (y-y')] \nonumber \\
&&\times  \sum_{m=-s}^{s-1} \sqrt{(3-m)(4+m)/4}\,\, \psi_m^{*} ({\bf r})\, \psi_{m+1} ({\bf r})  \,,
\nonumber \\
\label{Hd_11}
\end{eqnarray}
where $\gamma=g_L \mu_B $.

The dipolar matrix element $({\cal{H}}_{d})_{10}$ equals
\begin{eqnarray}
&&({\cal{H}}_d (\mathbf{r}))_{10} =  \nonumber \\
&&-\sqrt{3}\,  (3 \gamma^2) 
\int d^3r' \frac{[(x-x')-i(y-y')](z-z')}{|\mathbf{r}-\mathbf{r}'|^5} \nonumber \\
&& \times \sum_{m=-s}^s  m |\psi_m ({\bf r})|^2   \nonumber \\
&& -\sqrt{3}\,  (3 \gamma^2) 
\int d^3r' \frac{[(x-x')-i(y-y')]^2}{|\mathbf{r}-\mathbf{r}'|^5}  \nonumber \\
&&\times \sum_{m=-s+1}^s \sqrt{(4-m)(3+m)/4}\,\, \psi_m^{*} ({\bf r})\, \psi_{m-1} ({\bf r}) \nonumber \\
&& +\sqrt{3}\,  \gamma^2 
\int d^3r' \left[ \frac{2}{|\mathbf{r}-\mathbf{r}'|^3}-3\frac{(x-x')^2+(y-y')^2}
{|\mathbf{r}-\mathbf{r}'|^5} \right]   \nonumber \\
&& \times  \sum_{m=-s}^{s-1} \sqrt{(3-m)(4+m)/4}\,\, \psi_m^{*} ({\bf r})\, \psi_{m+1} ({\bf r})  \;.
\nonumber \\
\label{Hd_10}
\end{eqnarray}

\end{document}